\documentclass[3p,times,twocolumn]{elsarticle}

\usepackage{ecrc}
\usepackage{subfigure}
\usepackage[amssymb]{SIunits}

\volume{00}

\firstpage{1}

\journalname{Nuclear and Particle Physics Proceedings}

\runauth{J. Viinikainen}

\jid{nppp}

\jnltitlelogo{Nuclear and Particle Physics Proceedings}

\usepackage{amssymb}

\usepackage{lineno}

\usepackage[figuresright]{rotating}

\newcommand{\pp}{\ensuremath{\mathrm{p\kern-0.05em p}}}
\newcommand{\pPb}{\ensuremath{\mbox{p--Pb}}}
\newcommand{\sqrtS}{\ensuremath{\sqrt{s}}}
\newcommand{\sqrtSnn}{\ensuremath{\sqrt{s_{\mathrm{NN}}}}}
\newcommand{\sqrtSE}[2][TeV]{\ensuremath{\sqrtS = #2\,\mathrm{#1}}}
\newcommand{\sqrtSnnE}[2][TeV]{\ensuremath{\sqrtSnn = #2\,\mathrm{#1}}}
\newcommand{\GeVc}{\ensuremath{\mathrm{GeV}\kern-0.05em/\kern-0.02em c}}
\def\pt#1{\ensuremath{p_{\rm T#1}}} 
\def\jt#1{\ensuremath{j_{\rm T#1}}}
\def\vjt#1{\ensuremath{\vec{j}_{\rm T#1}}}
\newcommand{\xlong} {\ensuremath{x_{\parallel}}}
\def\la{\left< }
\def\ra{\right> }
\def\meankv#1{\ensuremath{\la#1^2\ra}}
\def\rms#1{\ensuremath{\sqrt{\meankv{#1}}}}
\def\fig#1{Figure~\ref{#1}}

\begin{document}

\begin{frontmatter}



\dochead{}

\title{Jet transverse fragmentation momentum from h--h correlations \\ in $\pp$ and $\pPb$ collisions}


\author{J.~Viinikainen for the ALICE collaboration}
\address{University of Jyv{\"a}skyl{\"a}}
\address{Helsinki Institute of Physics}


\begin{abstract}
QCD color coherence phenomena, like angular ordering, can be studied by looking at jet fragmentation. As the jet is fragmenting, it is expected to go through two different phases. First, there is QCD branching that is calculable in perturbative QCD. Next, the produced partons hadronize in a non-perturbative way later in a hadronization process. The jet fragmentation can be studied using the method of two particle correlations. A useful observable is the jet transverse fragmentation momentum $\jt{}$, which describes the angular width of the jet. In this contribution, a differential study will be presented in which separate $\jt{}$ components for branching and hadronization will be distinguished from the data measured by the ALICE experiment. The $\pt{t}$ dependence of the hadronization component $\rms{\jt{}}$ is found to be rather flat, which is consistent with universal hadronization assumption. However, the branching component shows slightly rising trend in $\pt{t}$. The $\sqrtSE{7}$ $\pp$ and $\sqrtSnnE{5.02}$ $\pPb$ data give the same results within error bars, suggesting that this observable is not affected by cold nuclear matter effects in $\pPb$ collisions. The measured data will also be compared to the results obtained from PYTHIA8 simulations.
\end{abstract}


\end{frontmatter}


\section{Introduction}
\label{sec:introduction}

\linespread{0.95}
\selectfont

\begin{figure}
    \begin{center}
      \includegraphics[width = 0.23\textwidth]{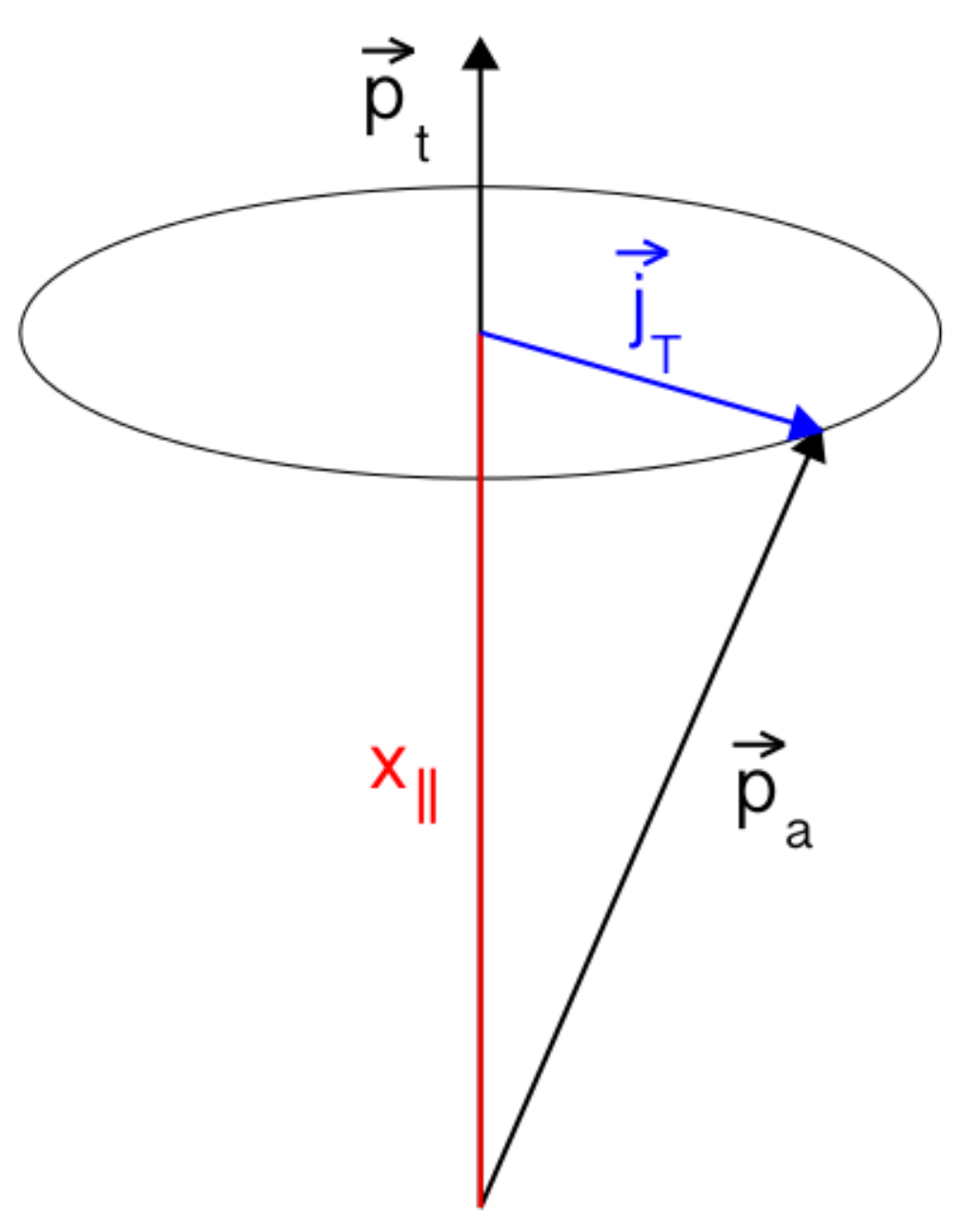}
    \end{center}
    \caption{Illustration of $\vjt{}$ and $\xlong$.  When the trigger particle is a high-$\pt{}$ particle that approximates the jet axis sufficiently well, the $\vjt{}$ can be written as the transverse momentum component of the associated particle momentum $\vec{p}_{\mathrm{a}}$ with respect to the trigger particle momentum $\vec{p}_{\mathrm{t}}$. The fragmentation variable $\xlong$ is the projection of $\vec{p}_{\mathrm{a}}$ to $\vec{p}_{\mathrm{t}}$ divided by $p_{\mathrm{t}}$.}
    \label{fig:jtdefinition}
  \end{figure}

In this work we study the jet transverse fragmentation momentum $\jt{}$, which is defined as the transverse momentum component of the jet fragment with respect to the jet axis. An illustration of $\jt{}$ is shown in \fig{fig:jtdefinition}. This quantity can be connected to two particle correlations by requiring that the trigger particle is the particle with the highest transverse momentum in an event (leading particle) and that this transverse momentum is sufficiently high. In this case the trigger particle momentum vector approximates the jet axis. Identifying the associated particle as the jet fragment, $\vjt{}$ becomes the transverse momentum component of the associated particle momentum $\vec{p}_{\mathrm{a}}$ with respect to the trigger particle momentum $\vec{p}_{\mathrm{t}}$. The length of the $\vjt{}$ vector is
  \begin{equation}
    \jt{} = \frac{|\vec{p}_{\mathrm{t}} \times \vec{p}_{\mathrm{a}}|}{|\vec{p}_{\mathrm{t}}|} \,.
  \label{eq:jtdefinition}
  \end{equation}
  In the analysis, the results are obtained as a function of the fragmentation variable $\xlong$. This is defined as the projection of the associated particle momentum to the trigger particle divided by the trigger particle momentum: 
  \begin{equation}
    \xlong = \frac{\vec{p}_{\mathrm{t}} \cdot \vec{p}_{\mathrm{a}}}{\vec{p}_{\mathrm{t}}^{2}} \,.
  \label{eq:xedefinition}
  \end{equation}
   Binning in $\xlong$ rather than $\pt{a}$ is chosen because $\xlong$ scales with the trigger $\pt{}$. We measure bins where the associated particles have similar momentum fraction relative to trigger. 
  
Because $\xlong$ follows the jet axis by construction, it is intuitive to define the near side with respect to this axis rather than using only azimuthal angle difference. The associated particle is defined to be in the near side if it is in the same hemisphere as the trigger particle:
  \begin{equation}
    \vec{p}_{\mathrm{t}} \cdot \vec{p}_{\mathrm{a}} > 0 \;.
    \label{eq:3Dnearside}
  \end{equation}
  
Previously $\jt{}$ has been measured for example by correlating the particles inside a jet cone with the reconstructed jet axis \cite{firstjtmeasurement,cdfreport,atlaksenJetit} or by calculating it from the azimuthal correlation function \cite{PHENIXjets}. Only one component for $\jt{}$ is extracted in these studies, describing the whole time evolution of the jet. In our study we want to be able to isolate different components for parton shower and hadronization. To do this, we follow a similar approach as is taken in PYTHIA \cite{lundString}. In this model jet fragmentation consists of two phases, a perturbative showering phase where partons lose their virtuality by emitting gluons followed by a non-perturbative hadronization phase where the partons combine to hadrons.

\begin{figure}
    \begin{center}
      \includegraphics[width = 0.34\textwidth]{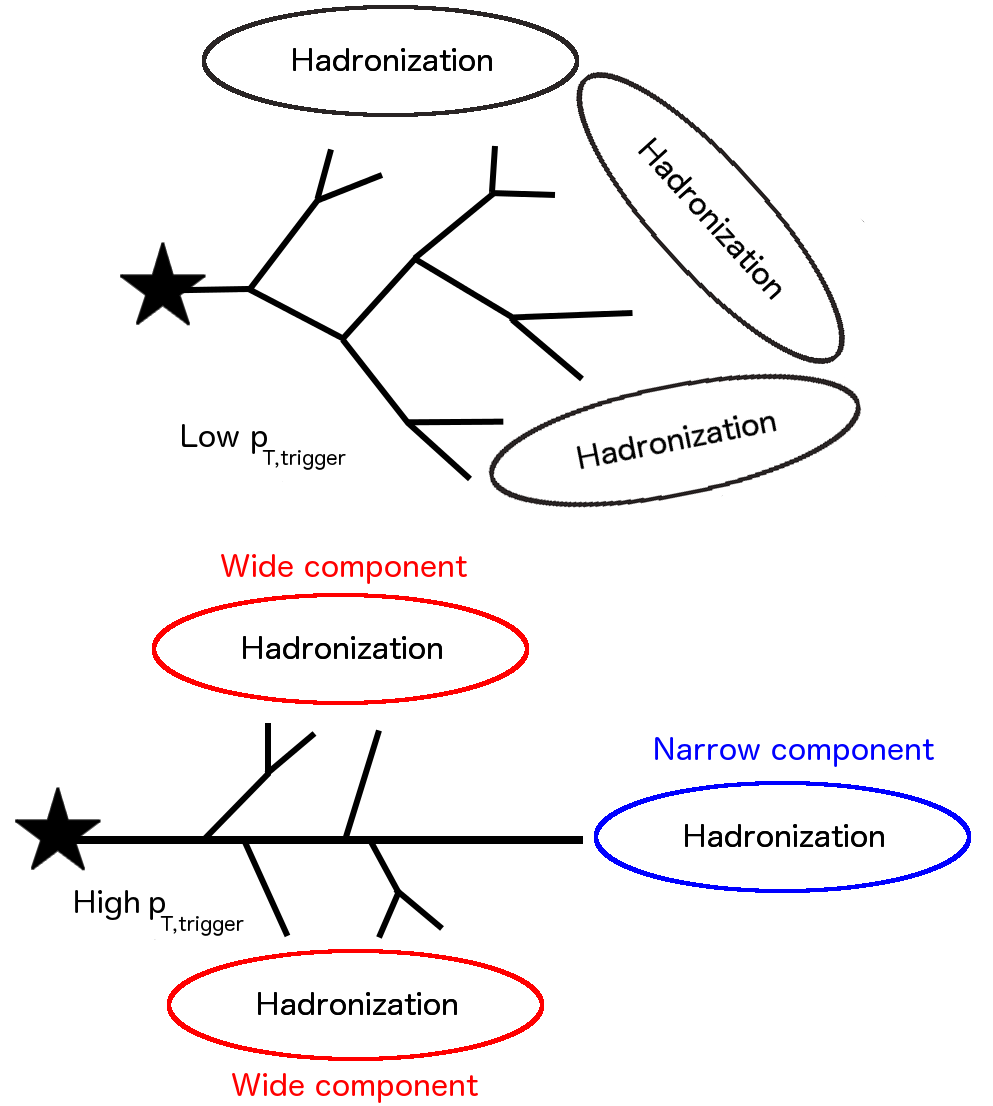}
    \end{center}
    \caption{\emph{Up:} If there is no high $\pt{}$ particle in the event, the showering and hadronization components of jet fragmentation are folded together. \emph{Down:} In the presence of a high $\pt{}$ particle, the showering and hadronization components can be separated from the data.}
    \label{fig:jetevolution}
  \end{figure}

Our working hypothesis is illustrated in \fig{fig:jetevolution}. If there are no sufficiently high $\pt{}$ particles in an event, the showering and hadronization components are folded together, as illustrated by the upper half of the figure. But if we restrict ourselves to events with a high $\pt{}$ leading particle, the folding is weaker and we are able to separate these two. Based on pQCD predictions like angular ordered parton cascades \cite{basicsofpqcd}, we expect that the leading parton emits soft gluons preferably to rather wide angles. Also the leading parton is not much affected by this soft emission. On the other hand, based on Lund string model \cite{lundString}, we assume that the non-perturbative hadronization produces particles in relatively narrow angles with respect to the hadronizing parton. Thus, if the leading particle is a good approximation for the leading parton, the particles near the leading particle are mainly coming from the hadronization of the leading parton. As the showering produces partons to wide angles and their direction is not changed much by the hadronization, the orientation of the produced particles with respect to the trigger particle is mainly determined by the showering phase. A two component fit to the $\jt{}$~distribution could allow us to separate the two phases.

\begin{figure}
    \begin{center}
      \includegraphics[width = 0.45\textwidth]{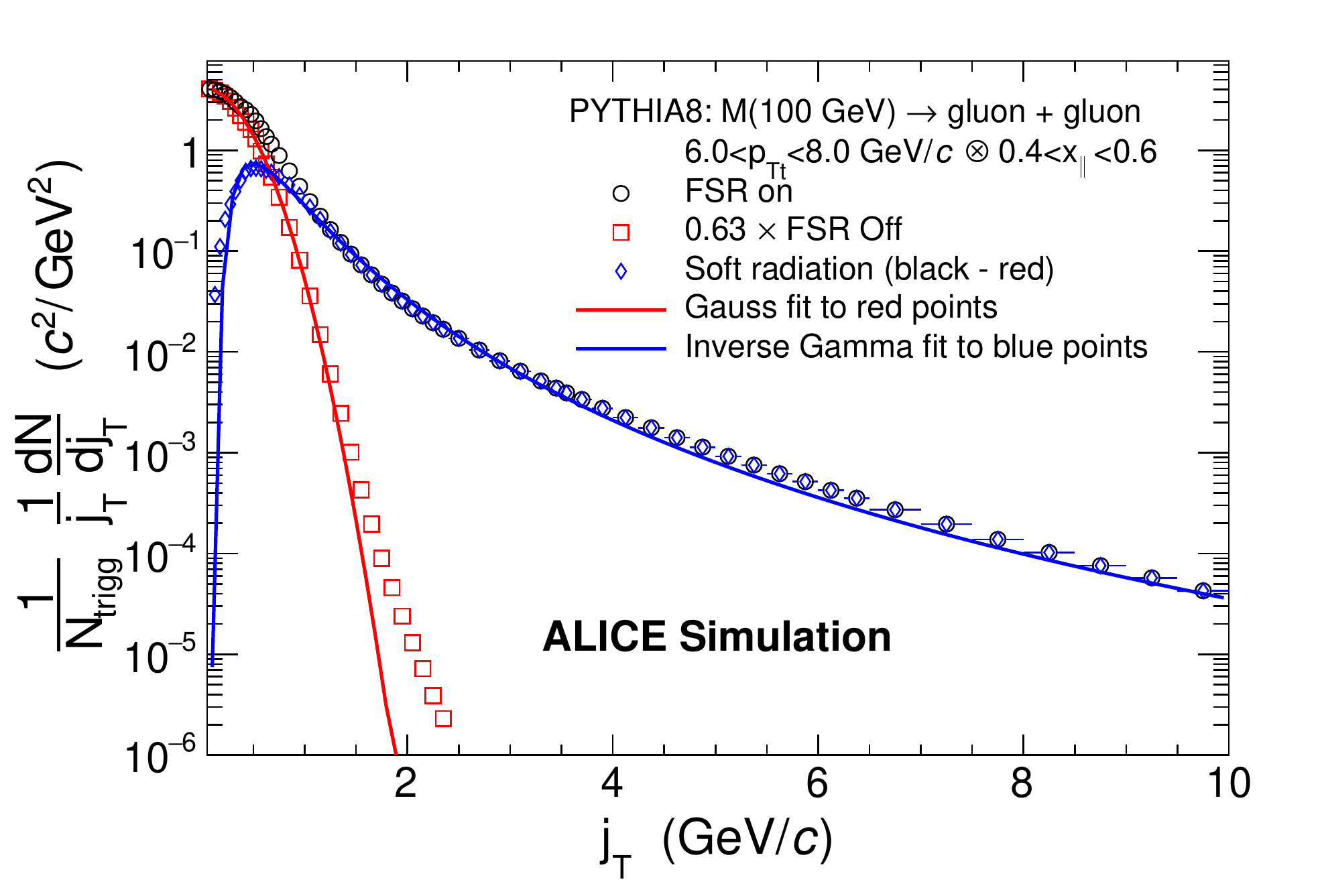}
    \end{center}
    \caption{Results from PYTHIA8 study with a di-gluon initial state. Black distribution is obtained when the final state radiation is on. When the final state radiation is turned off, the red distribution emerges. Blue one is calculated by subtracting the red from the black.}
    \label{fig:componentsFromResonance}
  \end{figure}

To see if the separation of two components is justifiable, a PYTHIA8 study was conducted. In this study a two gluon initial state was created to get as clean dijet samples as possible. Then the final state particles were produced controlling the presence of the final state radiation. Without final state radiation, the final state particles come purely from the hadronization of the leading parton. When the final state radiation is allowed, the partons go through both showering and hadronization before becoming final state particles. The results of this study are presented in \fig{fig:componentsFromResonance}. When the final state radiation is turned off, the component from hadronization appears as a narrow, Gaussian like distribution with a short tail. When the final state radiation is turned back on, a long tail appears after this peak. These observations support the idea behind the two component model. 

\section{Analysis methods}
\label{sec:methods}

\linespread{0.95}
\selectfont

The analysis is done using $\sqrtSE{7}$ $\pp$ ($3.0 \cdot 10^{8}$ events) and $\sqrtSnnE{5.02}$ $\pPb$ ($1.3 \cdot 10^{8}$ events) data recorded by the ALICE detector~\cite{Aamodt:2008zz}. The tracks are measured by the Inner Tracking System (ITS) and the Time Projection Chamber (TPC). One out of six possible hits is required in ITS and 70/159 in TPC. The innermost layer of ITS ($|\eta| < 2$) and the V0 detector ($2.8 < \eta < 5.1$ and $-3.7 < \eta < -1.7$) are used for triggering. Charged tracks with $\pt{} > 0.3\,\GeVc$ in TPC acceptance ($|\eta| < 0.8$) are selected for the analysis.

\begin{figure}
    \begin{center}
      \includegraphics[width = 0.32\textwidth]{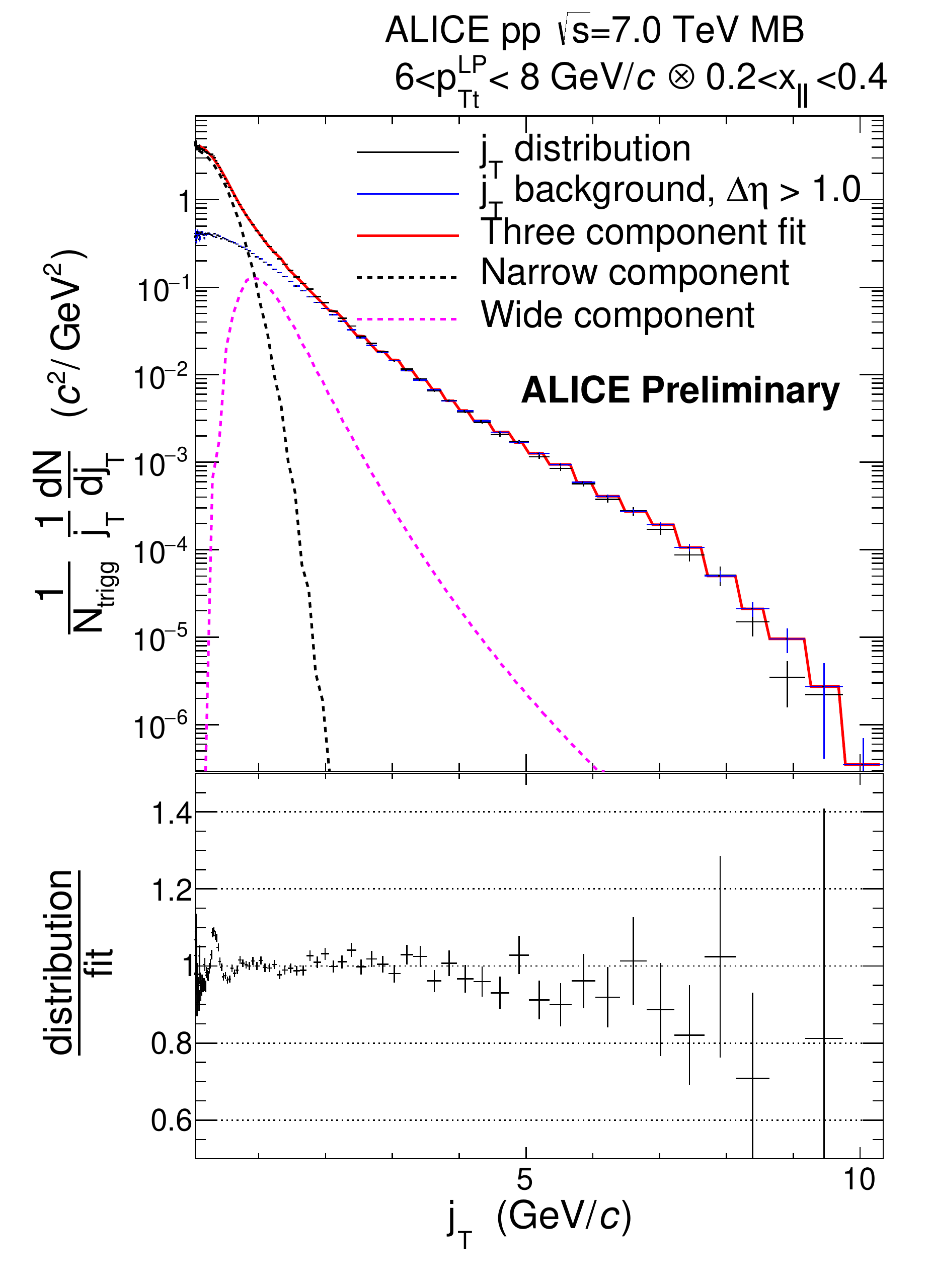}
    \end{center}
    \caption{Measured $\jt{}$ distribution with a three component fit into it. The three components describe the background (blue), hadronization (black dashed) and showering (magenta).}
    \label{fig:jtdistribution}
  \end{figure}

The chosen analysis method is two particle correlations. All charged particles inside each $\xlong$ bin are paired with a leading particle trigger and $\jt{}$ is calculated for each of these pairs. An example of a measured $\jt{}$ distribution is presented in \fig{fig:jtdistribution}. Here the black histogram contains both signal and background. The background is mostly coming from the underlying event. We use $\eta$-gap method to estimate the background contribution. The pairs with $\Delta\eta > 1.0$ are regarded as background pairs, since the jet correlation is expected to be a small angle correlation. We extrapolate the background to the signal region by generating new pairs from each background pair by randomizing $\eta$ for trigger and associated particles from inclusive $\eta$ distributions in corresponding $\pt{t}$ and $\pt{a}$ bins. This will give a background template for the analysis. This template, generated separately for each $\pt{t}$ and $\xlong$ bin, is then fitted to the $\jt{}$ distribution together with a Gaussian function for the narrow (hadronization) component and an Inverse Gamma function for the wide (showering) component. It can be seen from \fig{fig:jtdistribution} that the fit works very well, except in the region around $\jt{} \sim \unit{0.4}{GeV}$, where a small bump appears. Based on PYTHIA studies, this bump comes from events where the leading particle is a neutral meson decay product. When the other decay product is correlated with this trigger, the invariant mass of the pair brings an additional correlation component on top of the jet correlations. This is taken into account in the evaluation of systematic uncertainties. 

\section{Results}
\label{sec:results}

\linespread{0.94}
\selectfont

\begin{figure}[t]
    \vskip -4pt
    \begin{center}
     \subfigure{\includegraphics[width = 0.29\textwidth]{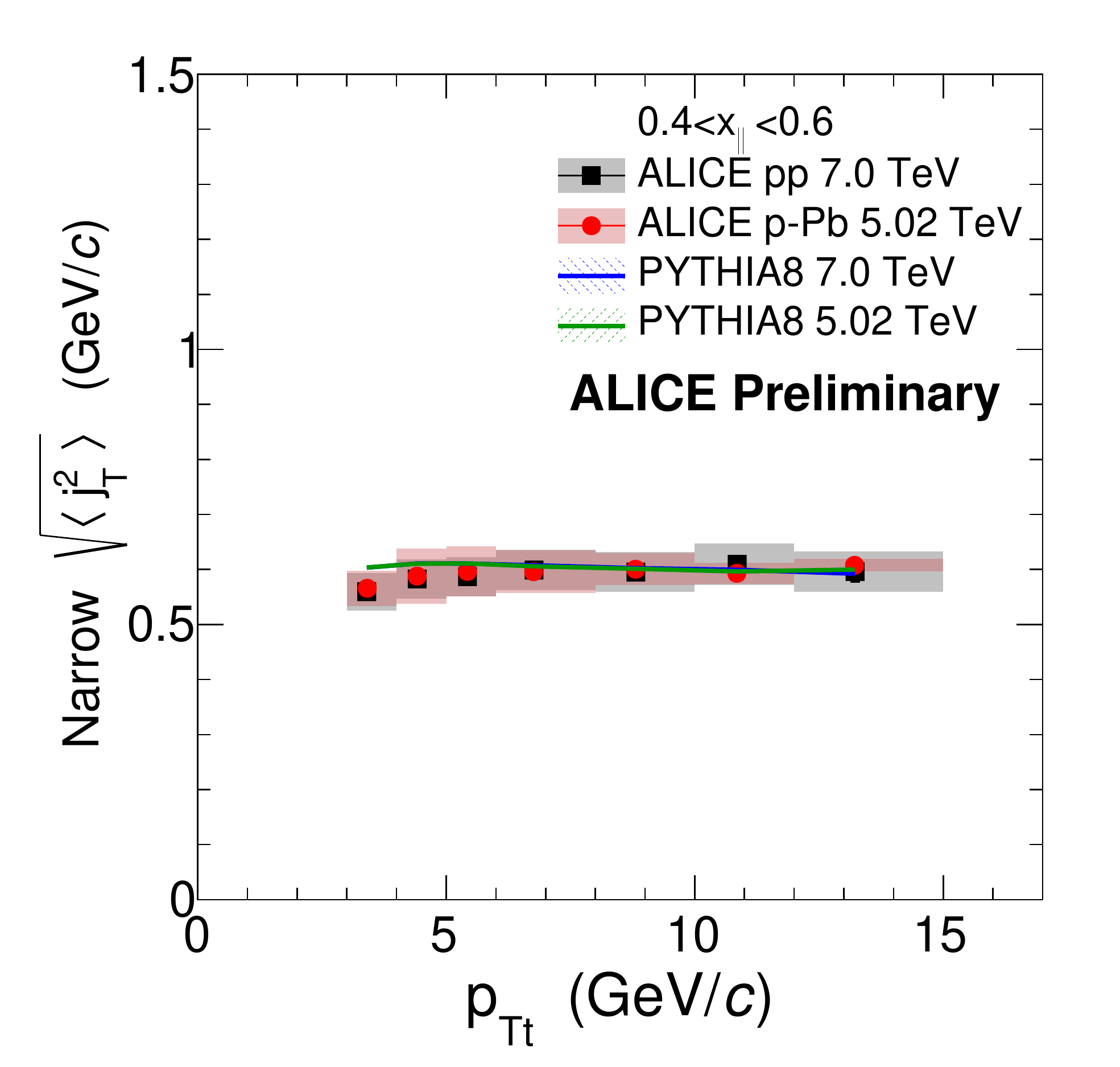}}
     \subfigure{\includegraphics[width = 0.29\textwidth]{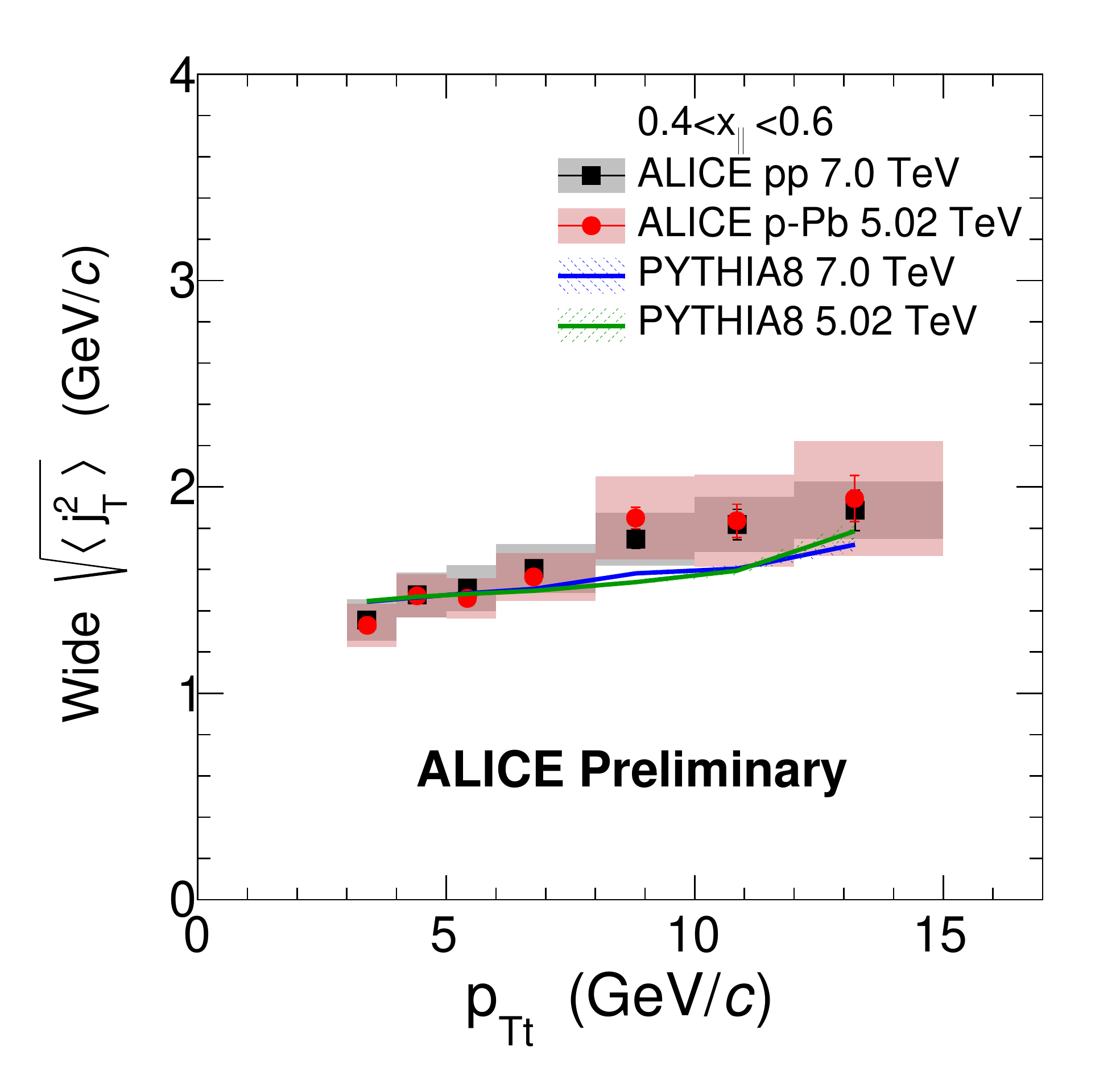}}
    \end{center}
    \caption{RMS values of narrow (hadronization) and wide (showering) $\jt{}$ components. Results from $\pp$ and $\pPb$ data agree very well together and with PYTHIA8.}
    \label{fig:jtrms}
  \end{figure}

The final $\rms{\jt{}}$ results are calculated from the obtained fit parameters for narrow and wide components. Based on earlier results \cite{firstjtmeasurement,PHENIXjets} the hadronization component is expected to be universal, meaning that it is $\sqrtS$ independent and that similar jets in different $\pt{t}$ (same $\xlong$~bin) give the same $\rms{\jt{}}$ results. It can be seen from \fig{fig:jtrms} that the narrow component results show a flat trend as a function of $\pt{t}$ and that there is no difference between results from $\pp$ and $\pPb$ data. Both of these observations support the universality expectation. Also PYTHIA8 seems to describe the data.

The completely new result in this work is the $\jt{}$ for the showering part of the jet fragmentation. This is the wide component in \fig{fig:jtrms}. Here we can see that there is a rising trend in $\pt{t}$. This can be explained by the fact that higher $\pt{}$ partons are likely to have higher virtuality. Higher virtuality leads to stronger gluon emission in the showering phase. When the emission is stronger, it is likely to increase the RMS of the distribution since more splittings are allowed in the shower. The fact that $\pp$ and $\pPb$ agree within the error bars suggests that there are no significant cold nuclear matter effects. Again, PYTHIA8 describes the data well.

\begin{figure}[t]
    \vskip -4pt
    \begin{center}
     \subfigure{\includegraphics[width = 0.29\textwidth]{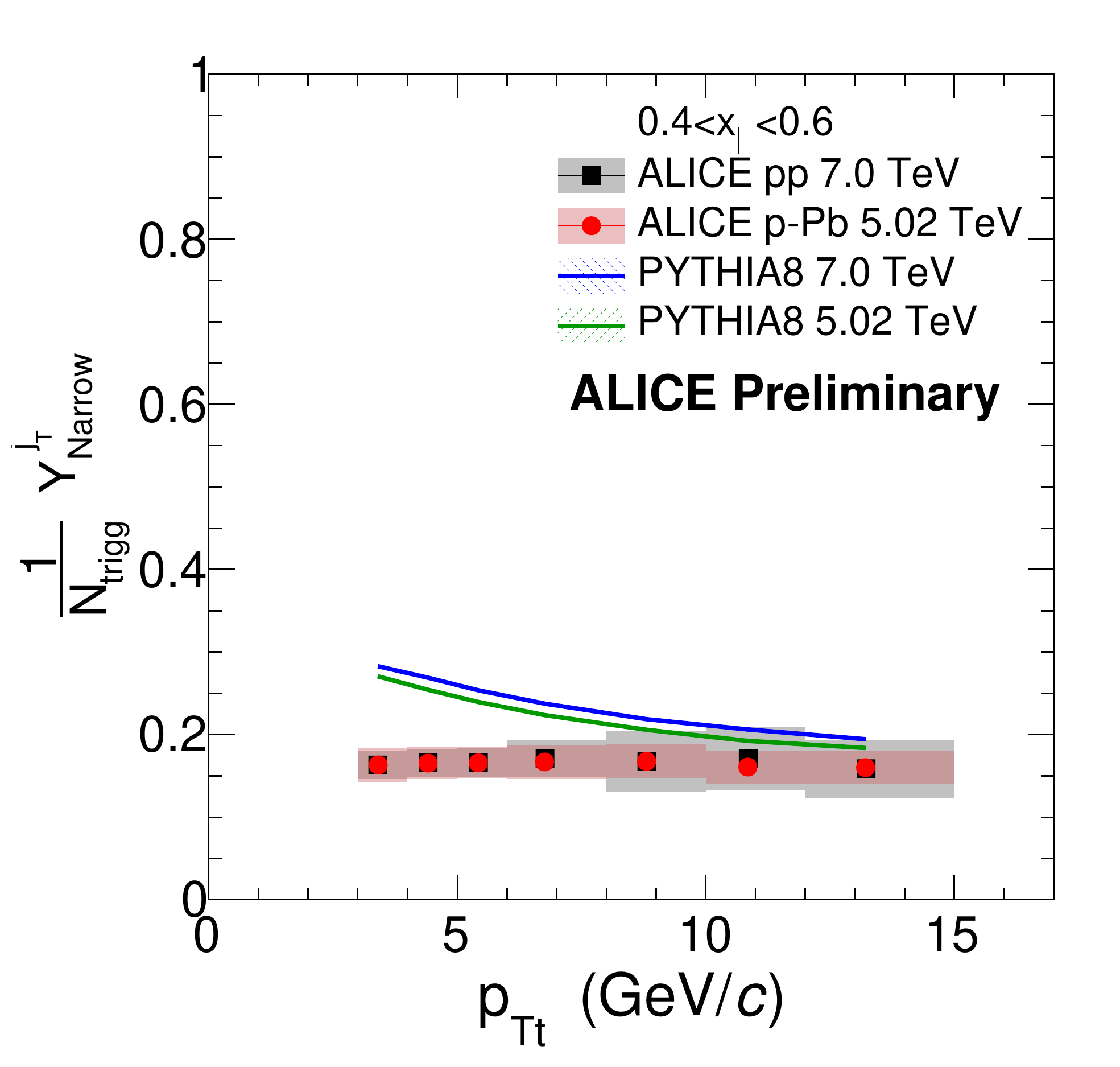}}
     \subfigure{\includegraphics[width = 0.29\textwidth]{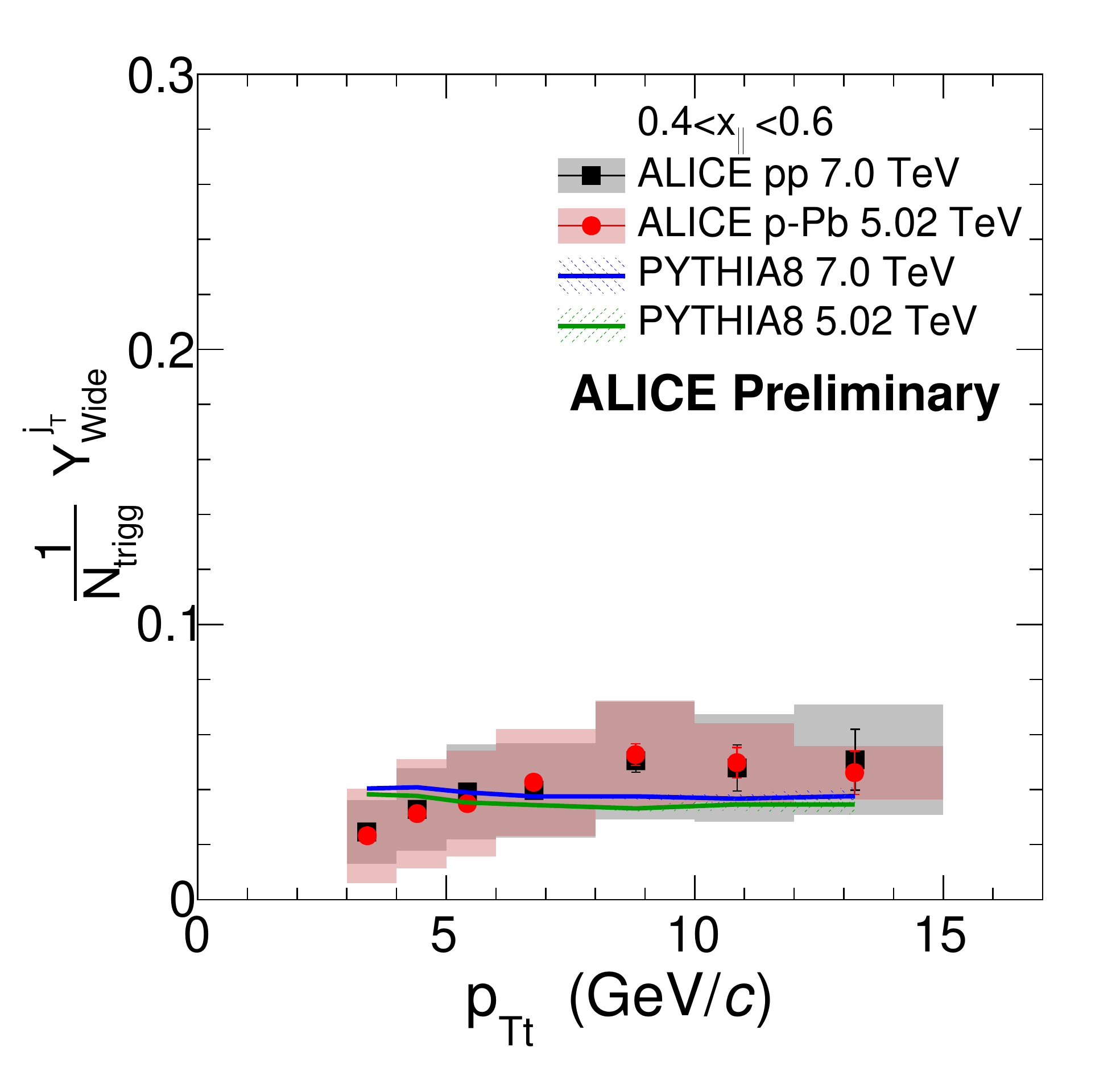}}
    \end{center}
    \caption{Yields of narrow (hadronization) and wide (showering) $\jt{}$ components. Results from $\pp$ and $\pPb$ data agree very well together but PYTHIA8 seems to overestimate the narrow component yields somewhat.}
    \label{fig:jtyield}
  \end{figure}

The final $\jt{}$ yield results are presented in \fig{fig:jtyield}. The narrow yield shows a flat trend in $\pt{t}$, the wide yield shows some hints of rising as a function of $\pt{t}$, but it could also be flat within current error bars. PYTHIA8 seems to be overestimating the yield of the narrow component, but manages to describe the wide component rather well. Similar behavior was earlier observed in an underlying event analysis in $\pp$ collisions at $\sqrtS = 0.9$ and $\unit{7}{TeV}$ \cite{ALICE:2011ac}.
  
\section{Conclusions}
\label{sec:conclusions}

\linespread{0.95}
\selectfont

Using two particle correlations, we extracted the RMS and yield for two different $\jt{}$ components, hadronization and showering. Our main observations are the following:

\begin{itemize}
  \item The narrow component RMS shows a flat $\pt{t}$ trend and the results between $\pp$ and $\pPb$ datasets agree with each other. This supports the universal hadronization expectation.
  \item The wide component RMS shows a rising $\pt{t}$ trend. This could be caused by increased parton emission allowed by increased leading parton virtuality.
  \item There is no difference in wide component RMS between $\pp$ and $\pPb$. No cold nuclear matter effects are seen.
  \item PYTHIA8 seems to describe all the RMS results and showering yield well. However, it overestimates the hadronization yield by up to 50 \%.
\end{itemize}





\linespread{1.00}
\selectfont

\nocite{*}
\bibliographystyle{elsarticle-num}
\bibliography{Viinikainen_J}







\end{document}